\documentclass[aps,twocolumn,superscriptaddress,showpacs]{revtex4}

\usepackage[dvips]{graphicx}  % for figure
\usepackage{color}  % for color  usage: \textcolor{red}{text}
\newcommand{\nc}{\newcommand}           % new command
\nc{\vc}[1]     {\mbox{\boldmath $#1$}} % boldmath(vector)
\nc{\bra}       {\langle}               % bra
\nc{\ket}       {\rangle}               % ket
\nc{\bras}[1]   {\langle#1|}            % <#1|
\nc{\kets}[1]   {|#1\rangle}            % |#1>
\nc{\del}       {\partial}              % bra state
\nc{\red}[1]    {\textcolor{red}{#1}}  % red
\nc{\blue}[1]   {\textcolor{blue}{#1}}  % red
\nc{\green}[1]   {\textcolor{green}{#1}}  % green

\nc{\mydraft}	{\setlength{\topmargin}{-1.0cm}} % move upside
\mydraft

\begin{document}

\title{Role of tensor interaction in He isotopes with tensor-optimized shell model}

\author{Takayuki Myo\footnote{myo@ge.oit.ac.jp}}
\affiliation{General Education, Faculty of Engineering, Osaka Institute of Technology, Osaka, Osaka 535-8585, Japan}
\affiliation{Research Center for Nuclear Physics (RCNP), Osaka University, Ibaraki, Osaka 567-0047, Japan}

\author{Atsushi Umeya\footnote{aumeya@nit.ac.jp}}
\affiliation{Human Science and Common Education, Faculty of Engineering, Nippon Institute of Technology, Saitama 345-8501, Japan}

\author{Hiroshi Toki\footnote{toki@rcnp.osaka-u.ac.jp}}
\affiliation{Research Center for Nuclear Physics (RCNP), Osaka University, Ibaraki, Osaka 567-0047, Japan}

\author{Kiyomi Ikeda\footnote{k-ikeda@postman.riken.go.jp}}
\affiliation{RIKEN Nishina Center, Wako, Saitama 351-0198, Japan}

\date{\today}

\begin{abstract}
We studied the role of the tensor interaction in He isotopes systematically on the basis of the tensor-optimized shell model (TOSM).  We use a bare nucleon-nucleon interaction AV8$^\prime$ obtained from nucleon-nucleon scattering data.  The short-range correlation is treated in the unitary correlation operator method (UCOM). Using the TOSM+UCOM approach, we investigate the role of tensor interaction on each spectrum in He isotopes.  It is found that the tensor interaction enhances the $LS$ splitting energy observed in $^5$He, in which the $p_{1/2}$ and $p_{3/2}$ orbits play different roles on the tensor correlation. In $^{6,7,8}$He, the low-lying states containing extra neutrons in the $p_{3/2}$ orbit gain the tensor contribution. On the other hand, the excited states containing extra neutrons in the $p_{1/2}$ orbit lose the tensor contribution due to the Pauli-blocking effect with the $2p2h$ states in the $^4$He core configuration.
\end{abstract}


%\keywords{
%neutron halo; neutron skin, complex scaling method 
%}

\maketitle 

\section{Introduction}

It is an important subject in nuclear physics to understand the nuclear structure from the viewpoint of the nucleon-nucleon ($NN$) interaction.  The $NN$ interaction has distinctive features. There exist strong tensor interactions at long and intermediate distances caused by pion exchange and strong central repulsions at a short distance caused by quark dynamics. It is important to investigate the nuclear structure in relation to the above characteristics of the $NN$ interaction~\cite{akaishi86,kamada01}.  Recently, it has become possible to calculate nuclei up to mass around $A\sim 12$ using a $NN$ interaction with the Green's Function Monte Carlo method~(GFMC) \cite{pieper01, pudliner97}.  It is, however, extremely time consuming to apply this method to heavier nuclei.  It is desired to develop a new method to calculate nuclear structure with large nucleon numbers by taking care of the characteristic features of the $NN$ interaction.

The presence of the tensor interaction in the $NN$ interaction causes an explicit $d$-wave component in the relative wave function in a nucleus, in particular, for the proton-neutron ($pn$) pair as seen in the deuteron.  The $d$-wave component in the deuteron is essential to bind the system via the $sd$ coupling of the tensor interaction.  This $d$-wave component is found to be spatially compact as compared with the $s$-wave component due to the large momentum components brought by the tensor interaction~\cite{ikeda10}.  This effect originates from the pseudo-scalar nature of the one-pion exchange.  We mention also that a large fraction of the $pn$ pair is observed experimentally than $pp$ or $nn$ pairs in light nuclei \cite{subedi08, simpson11}.  This enhancement of the $pn$ pair is hard to reproduce theoretically in a simple shell model (mean-field picture) \cite{simpson11} except for the rigorous method as the one of GFMC \cite{schiavilla07}, which treats the tensor interaction explicitly. It is important to make efforts to study the high momentum components caused by the tensor interaction in finite nuclei \cite{tanihata10}.

There are two important developments to proceed nuclear structure calculations to heavy nuclei and to see the dynamics induced by the bare $NN$ interaction. One is to find out that the strong tensor interaction is of intermediate range and we are able to express the tensor correlation in a reasonable shell model space~\cite{myo07,myo09}.  We name this method as Tensor Optimized Shell Model (TOSM).  The other is the Unitary Correlation Operator Method (UCOM) to treat the short-range correlation caused by the short-range repulsion~\cite{feldmeier98, neff03, roth10}. We shall combine two methods, TOSM and UCOM, to describe nuclei using bare $NN$ interaction.  In the TOSM part, the wave function is constructed in terms of the shell model basis states with full optimization of the two particle-two hole ($2p2h$) states. This means that there is no truncation of the particle states within the $2p2h$ space of TOSM. In particular, the spatial shrinkage of the particle states is essential to obtain convergence of the tensor contribution involving the high momentum components~\cite{shimizu74,toki02,sugimoto04,ogawa06}, which is also seen in the deuteron. This treatment of the bare tensor interaction in TOSM corresponds to the one-pair approximation correlated by the tensor interaction~\cite{togashi07, ogawa06}.  In a few-body viewpoint, Horii {\it et al.} have already confirmed the reliability of one-pair approximation to treat the bare tensor interaction using the few-body method for $s$-shell nuclei by reproducing more than 90\% of the tensor correlation energy~\cite{horii11}. The explicit inclusion of the $2p2h$ states in the extended mean field model for heavy nuclei has been formulated by Ogawa {\it et al.} \cite{ogawa11}.

So far, we have obtained successful results using TOSM for the investigation of the tensor correlations in He and Li isotopes.
In $^4$He, we have shown the selectivity of $(p_{1/2})^2(s_{1/2})^{-2}$ of $2p2h$ state with the $pn$ pair induced by the tensor interaction and have recognized this correlation as deuteron-like one \cite{myo09}.  In $^{5}$He, this $pn$ tensor correlation in $^4$He blocks the $p_{1/2}$ occupation of a last neutron by the Pauli-principle \cite{myo05}. This blocking occurred in $1/2^-$ state of $^5$He produces the $p$-wave splitting energy in $^5$He.  In $^{10}$Li and $^{11}$Li, we have performed the coupled two- and three-body analyses with the $^9$Li core described in TOSM, respectively.  Similar to the case of $^5$He, the $p_{1/2}$ occupation of last one and two neutrons are blocked by the tensor correlation in $^9$Li.  As a result, we have naturally explained the virtual $s$-state in $^{10}$Li, and the large $s$-wave mixing and a neutron halo formation in $^{11}$Li.
We have also reproduced the various observables in $^{10}$Li and $^{11}$Li~\cite{myo07_11, myo08}.

For the structures of He isotopes, many experiments have been reported in neutron-rich side \cite{skaza06, skaza07, golovkov09}. However, there are still contradictions in the observed energy levels and their spin assignments. We have performed the analyses of He isotopes on the basis of the cluster model treating the continuum states for extra neutrons \cite{myo07_7, myo09_7, myo10}.  We have reproduced various experimental data and predicted energy levels and decay widths.  In these analyses, the $^4$He core is assumed to be the $(0s)^4$ configuration and the effects of $2p2h$ excitations are not considered explicitly. In this approach, the dynamical coupling between the strong $2p2h$ excitations induced by the tensor interaction in $^4$He and the motions of extra neutrons in neutron-rich He isotopes cannot be treated. It is an interesting problem to take into account this coupling explicitly and to see how the coupling affects the energy levels and their tensor contributions in He isotopes.

In this paper, we proceed with our study of the TOSM+UCOM approach to neutron-rich He isotopes and discuss their structures focusing on the roles of the tensor interaction on the energies and configurations.  In the previous studies of $^{5,6}$He~\cite{myo05, myo06}, we have discussed the Pauli-blocking effect on the spectra of $^{5,6}$He based on the $^4$He+$n$+$n$ three-body model, in which the $^4$He core is described in the TOSM. Semi-microscopic potentials between $^4$He and an extra neutron was used, in which the tensor correlation is renormalized into the potential.  In those analyses, the importance of the $(p_{1/2})^2$ configuration of a $pn$ pair in the $^4$He core is shown to produce the splitting energies between the states corresponding to the $LS$-partners in $^{5,6}$He.  In the present study, we perform a full microscopic analysis by using the TOSM+UCOM approach with the bare $NN$ interaction for He isotopes and see how the present method works.  We discuss the role of the $p_{1/2}$ orbit excited by the tensor interaction and its structure dependence in the individual states obtained in $^{5-8}$He. We also see the kinetic energy behavior and the high momentum component in each state in relation with the tensor correlation. This subject becomes a foundation of the nuclear structure study on the basis of TOSM+UCOM.

In Sec.~\ref{sec:model}, we explain the method of the TOSM+UCOM approach for He isotopes.  In Sec.~\ref{sec:result}, we show the level structures of He isotopes and discuss their characteristics in relation with the tensor interaction.  A summary is given in Sec.~\ref{sec:summary}.

%%%%%%%%%%%%%%%%%%%%%%%%%%%%%%%%%%%%%%%%%%%%%%%%%%%%%%%%%%%%%
\section{Model}\label{sec:model}

\subsection{Tensor-optimized shell model (TOSM) for He isotopes}

We explain the framework of TOSM. We shall begin with a many-body Hamiltonian,
\begin{eqnarray}
    H
&=& \sum_i T_i - T_{\rm c.m.} + \sum_{i<j} V_{ij} , 
    \label{eq:Ham}
    \\
    V_{ij}
&=& v_{ij}^C + v_{ij}^{T} + v_{ij}^{LS} + v_{ij}^{Clmb} .
\end{eqnarray}
Here, $T_i$ is the kinetic energy of each nucleon with $T_{\rm c.m.}$ being the center of mass kinetic energy.  We take a bare interaction $V_{ij}$ such as AV8$^\prime$ \cite{pudliner97} consisting of central $v^C_{ij}$, tensor $v^T_{ij}$ and spin-orbit $v^{LS}_{ij}$ terms and $v_{ij}^{Clmb}$ the Coulomb term.  We obtain the many-body wave function $\Psi$ with the Schr\"odinger equation $H \Psi=E \Psi$.

We give the TOSM wave function $\Psi$ as 
\begin{eqnarray}
\Psi&=& \sum_{k_0} A_{k_0} \kets{0p0h;k_0} + \sum_{k_1} A_{k_1} \kets{1p1h;k_1}
\nonumber\\
&+& \sum_{k_2} A_{k_2} \kets{2p2h;k_2}  ,
      \label{eq:config}
\end{eqnarray}
\begin{eqnarray}
    \kets {1p1h;k_1}
&=& \bigl| \psi^{n_1}_{\alpha_1}(\vc{r}_1) \times \tilde\psi^{n_2}_{\alpha_1}(\vc{r}_1) \bigr\rangle ,
      \label{eq:1p1h}
    \\
    \kets {2p2h;k_2}
&=& \bigl| \psi^{n_1}_{\alpha_1}(\vc{r}_1) \psi^{n_2}_{\alpha_2}(\vc{r}_2)
    \times \tilde\psi^{n_3}_{\alpha_3}(\vc{r}_1) \tilde\psi^{n_4}_{\alpha_4}(\vc{r}_2) \bigr\rangle .
      \label{eq:2p2h}
\end{eqnarray}
Here, we omit the symbols of angular momentum coupling and total isospin for simplicity.  The states $\kets{0p0h;k_0}$ are the $0p0h$ shell model states. The configurations $\kets{1p1h;k_1}$ and $\kets{2p2h;k_2}$ are the $1p1h$ and $2p2h$ states with various radial components for particle states.  The vectors $\vc{r}_1$ and $\vc{r}_2$ represent the positions of nucleons. The labels $k_0$, $k_1$ and $k_2$ are the representative indices to distinguish various configurations.  We take the available configurations with the fixed spin and parity $J^\pi$ and the isospin of the total wave function $\Psi$. The amplitudes $\{A_{k_0},A_{k_1},A_{k_2}\}$ of the configurations are variational coefficients. The basis functions $\psi^n_{\alpha}$ and $\tilde\psi^n_{\alpha}$ are to describe particle and hole states of one nucleon with quantum numbers $n$ and $\alpha$, respectively; the index $n$ is to distinguish different radial basis functions and $\alpha$ are the sets of the spin-isospin quantum numbers to distinguish the orbits.  In TOSM, the hole states are described by harmonic oscillator basis states, whose length parameters are determined independently for each orbit in each $J^\pi$ state so as to minimize the total energy of the system.  For $^4$He, the $0p0h$ configuration is given by the $(0s)^4$ wave function and for $^{5-8}$He, the $0p0h$ configurations are given in the $0s+0p$ space with $0\hbar\omega$ excitations. The available configurations for particle states are taken into account within the $1p1h$ and $2p2h$ states with up to the partial wave $L_{\rm max}$ of $\psi^n_{\alpha}$, which determines the model space of TOSM.  We check the convergence of the solutions by increasing $L_{\rm max}$.  The $2p2h$ states play an important role on the description of the strong tensor correlation in TOSM and the $1p1h$ states can improve the radial properties of the hole states.

We employ the Gaussian expansion method to describe single-particle basis states $\psi^n_\alpha$ for particle states in TOSM~\cite{hiyama03, aoyama06}.  Each Gaussian basis function has the form of a nodeless harmonic oscillator wave functions (HOWF), except for the $1s$ and $1p$ orbits.  When we superpose a sufficient number of Gaussian basis functions with various length parameters, the radial component of the particle states can be fully optimized in every configuration of TOSM with respect to the Hamiltonian in Eq. (\ref{eq:Ham}).  The basis states for particle states should be orthogonal to the hole states having HOWF. This condition is imposed by using the Gram-Schmidt orthonormalization which is explained later. Technically, in order to use the non-orthogonal Gaussian basis function in the shell model framework, we construct the following orthonormalized single-particle basis function $\psi^n_{\alpha}$ used in Eqs.~(\ref{eq:1p1h}) and (\ref{eq:2p2h}) by using a linear combination of Gaussian bases $\{\phi_\alpha\}$ with length parameter $b_{\alpha,\nu}$.
\begin{eqnarray}
        \psi^n_{\alpha}(\vc{r})
&=&     \sum_{\nu=1}^{N_\alpha} d^n_{\alpha,\nu}\ \phi_{\alpha}(\vc{r},b_{\alpha,\nu}),
        \label{eq:Gauss1}
        \\
        \bra \psi^n_{\alpha} | \psi^{n'}_{\alpha'}\ket
&=&     \delta_{n,n'}\delta_{\alpha,\alpha'},
        \label{eq:Gauss3}
	\\
{\rm for}~~n~&=&~1,\cdots,N_\alpha,       
        \nonumber
\end{eqnarray}
where $N_\alpha$ is a number of basis functions for the orbit $\alpha$, and $\nu$ is an index that distinguishes the bases with different values of $b_{\alpha,\nu}$.
The explicit form of the Gaussian basis function is expressed as
\begin{eqnarray}
        \phi_{\alpha}(\vc{r},b_{\alpha,\nu})
&=&     N_l(b_{\alpha,\nu}) r^l e^{-(r/b_{\alpha,\nu})^2/2} [Y_{l}(\hat{\vc{r}}),\chi^\sigma_{1/2}]_j \chi_{t_z},
        \label{eq:Gauss2}
        \\
        N_l(b_{\alpha,\nu})
&=&     \left[  \frac{2\ b_{\alpha,\nu}^{-(2l+3)} }{ \Gamma(l+3/2)}\right]^{\frac12},
\end{eqnarray}
where $l$ and $j$ are the orbital and total angular momenta of the basis states, respectively, and $t_z$ is the projection of the nucleon isospin. The weight coefficients $\{d^n_{\alpha,\nu}\}$ are determined to satisfy the overlap condition in Eq.~(\ref{eq:Gauss3}). This is done by solving the eigenvalue problem of the norm matrix of the Gaussian basis set in Eq.~(\ref{eq:Gauss2}) with the dimension $N_\alpha$. Another method is the Gram-Schmidt orthonormalization, in which the basis function $\psi^n_{\alpha}$ is expanded by the Gaussian basis functions with the number $n$ \cite{myo07}. These two methods give the different $\{d^n_{\alpha,\nu}\}$ to prepare $\{\psi^n_{\alpha}\}$, but, are equivalent to obtaining the variational solutions of the total wave function $\Psi$ in Eq.~(\ref{eq:config}). This property is satisfied for any Gaussian basis number $N_\alpha$. Following this procedure, we obtain the new single-particle basis states $\{\psi^n_{\alpha}\}$ in Eq.~(\ref{eq:Gauss1}) used in  TOSM. The particle states in each configuration are determined independently in TOSM from the variational principle.  

We should construct Gaussian basis functions of particle states to be orthogonal to the occupied states, the $0s$ and the $0p$ orbits in He isotopes. We explain this procedure as follows; For the $0s$ orbit in the hole state, we employ one Gaussian basis function, namely, HOWF with length $b_{0s,\nu=1}$ = $b_{0s}$. For the $1s_{1/2}$ basis states in particle states, we introduce an extended $1s$ basis function orthogonal to the $0s_{1/2}$ state and that possesses a length parameter $b_{1s,\nu}$ that can differ from $b_{0s}$. In the extended $1s$ basis functions, the polynomial part is changed from the usual $1s$ basis states to satisfy the conditions of the normalization and the orthogonality to the $0s$ state \cite{myo07}. For the $1p$ states, we take the same method as used for the $1s$ case. One-body and two-body matrix elements in the Hamiltonian are analytically calculated using the Gaussian bases, whose explicit forms are given in appendix of Ref. \cite{myo09} for central, $LS$ and tensor interactions, respectively. In the numerical calculation, we prepare 10 Gaussian basis functions at most with various range parameters to get a convergence of the energy and Hamiltonian components.

We furthermore take care of the center-of-mass excitations. Toward this end, we take the well-tested method of introducing a Hamiltonian of center-of-mass motion in the many-body Hamiltonian known as the Lawson method~\cite{lawson}.  In the present study, we take the value of $\hbar \omega$ for the center of mass motion as the averaged one used in the $0s$ and $0p$ orbits in the $0p0h$ states with the weight of the occupation numbers in each orbit.  We have checked that this choice of $\hbar\omega$ gives variationally better solutions in the present scheme.  Adding this center of mass Hamiltonian as the Lagrange multiplier to the original Hamiltonian in Eq.~(\ref{eq:Ham}), we can effectively project out only the lowest HO state for the center-of-mass motion.  The value of the Lagrange multiplier is taken typically as 10-20 to suppress the center of mass excitations below 100 keV.

The variation of the energy expectation value with respect to the total wave function $\Psi$ in Eq.~(\ref{eq:config}) is given by
\begin{eqnarray}
\delta\frac{\bra\Psi|H|\Psi\ket}{\bra\Psi|\Psi\ket}&=&0\ ,
\end{eqnarray}
which leads to the following equations:
\begin{eqnarray}
    \frac{\del \bra\Psi| H - E |\Psi \ket} {\del b_{\alpha,\nu}}
&=& 0\ ,\quad
   \label{eq:vari1}
    \\
    \frac{\del \bra\Psi| H - E |\Psi \ket} {\del A_{k_i}}
&=&  0\qquad \mbox{for}~i=0,1,2 .
   \label{eq:vari2}
\end{eqnarray}
Here, total energy is represented by $E$.
The parameters $\{b_{\alpha,\nu}\}$ for the Gaussian bases appear in non-linear forms in the energy expectation value. 
We solve two kinds of variational equations in Eqs. (\ref{eq:vari1}) and (\ref{eq:vari2}) in the following steps.  
First, fixing all the length parameters $b_{\alpha,\nu}$ and the partial waves of basis states up to $L_{\rm max}$, 
we solve the linear equation for $\{A_{k_i}\}$ as an eigenvalue problem for $H$. 
We thereby obtain the eigenvalue $E$, which is functions of $\{b_{\alpha,\nu}\}$ and $L_{\rm max}$. 
Next, we try to adopt various sets of the length parameters $\{b_{\alpha,\nu}\}$ and expand the configuration space of TOSM by increasing $L_{\rm max}$ in order to find the solution which minimizes the total energy $E$. 
In the TOSM wave function, we can describe the spatial shrinkage of particle states with an appropriate radial form
in each configuration within $2p2h$ space, which is important to describe the tensor correlation \cite{myo07}.

%%%%%%%%%%%%%%%%%%%%%%%%%%%%%%%%%%%%%%%%%%%%%%
\subsection{Unitary Correlation Operator Method (UCOM)}
We explain UCOM for the short-range central correlation \cite{feldmeier98,neff03,roth10}, 
in which the following unitary operator $C$ is introduced
\begin{eqnarray}
C     &=&\exp(-i\sum_{i<j} g_{ij})~.
\label{eq:ucom}
\end{eqnarray}
We express the correlated wave function $\Psi$ in terms of less sophisticated wave function $\Phi$ as $\Psi=C\Phi$.  Hence, the Schr\"odinger equation becomes $\hat H \Phi=E\Phi$ where the transformed Hamiltonian is given as $\hat H=C^\dagger H C$.   Since the operator $C$ is expressed with a two-body operator in the exponential, it is in principle a many-body operator.  In case of the short-range correlation, we are able to truncate the modified operators at the level of two-body operators~\cite{feldmeier98}.

Two-body Hermite operator $g$ in Eq.~(\ref{eq:ucom}) is defined as
\begin{eqnarray}
g &=& \frac12 \left\{ p_r s(r)+s(r)p_r\right\} ~,
\label{eq:ucom_g}
\end{eqnarray}
\begin{eqnarray}
\frac{dR_+(r)}{dr}&=& \frac{s \left( R_+(r) \right)}{s(r)} ~, 
\end{eqnarray}
where the operator $p_r$ is the radial component of the relative momentum and is conjugate to the relative coordinate $r$. 
The function $s(r)$ expresses the amount of the shift of the relative wave function which depends on the relative coordinate.  In UCOM, we use $R_+(r)$ instead of $s(r)$, where the function $R_+(r)$ corresponds to the transformed relative coordinate from the original $r$. This $R_+(r)$ represents the correlation function to reduce the short-range amplitude of the relative wave functions so as to avoid the short-range repulsion in the $NN$ interaction. The explicit transformation of the operator is given in Refs. \cite{feldmeier98,neff03}.  We use the TOSM basis states to describe $\Phi$ which includes the tensor correlation.

In UCOM, the shift operator $g$ in Eq.~(\ref{eq:ucom_g}) is introduced for every nucleon pair in nuclei. The amount of the shifts, namely the the function $R_+(r)$ is determined variationally. We parametrize $R_+(r)$ in the same manner as proposed by Feldmeier and Neff \cite{feldmeier98,neff03}.  We assume the following forms for even and odd channels, respectively.
\begin{eqnarray}
	R_+^{\rm even}(r)
&=&	r + \alpha \left(\frac{r}{\beta}\right)^\gamma \exp[-\exp(\frac{r}{\beta})] , 
        \label{eq:R+}
	\\
	R_+^{\rm odd}(r)
&=&	r + \alpha \left( 1- \exp(-\frac{r}{\gamma}) \right) \exp[-\exp(\frac{r}{\beta})] .
        \label{eq:R-}
\end{eqnarray}
Here, $\alpha$, $\beta$, $\gamma$ are the variational parameters to optimize the $R_+(r)$ functions and minimize the total energy of the system.  They are independently determined for four channels of spin-isospin pair.  In the actual procedure of the variation, once we fix the parameters in $R_+(r)$, we solve the eigenvalue problem of the transformed Hamiltonian using Eqs. (\ref{eq:vari1}) and (\ref{eq:vari2}) and determine the configuration mixing in TOSM.  Next, we try to search various sets of the $R_+(r)$ parameters to minimize the total energy.

%%%%%%%%%%%%%%%%%%%%%%%%%%%%%%
\begin{table}[t]
\begin{center}
\caption{Optimized parameters in $R_+(r)$ used for TOSM+UCOM in four spin-isospin channels.}
\label{tab:R+2} 
\begin{tabular}{c|cccc}
\noalign{\hrule height 0.5pt}
              & $\alpha$ &  $\beta$ & $\gamma$ \\
\noalign{\hrule height 0.5pt}
singlet even  &~1.32~    &~ 0.88 ~  & ~ 0.36~ \\
triplet even  &~1.33~    &~ 0.93 ~  & ~ 0.41~ \\
singlet odd   &~1.57~    &~ 1.26 ~  & ~ 0.73~ \\
triplet odd   &~1.18~    &~ 1.39 ~  & ~ 0.53~ \\
\noalign{\hrule height 0.5pt}
\end{tabular}
\end{center}
\end{table}
%%%%%%%%%%%%%%%%%%%%%%%%%%%%%%
%%%%%%%%%%%%%%%%%%%%%%%%%%%%%%
\begin{figure}[t]
\begin{center}
\includegraphics[width=7.0cm,clip]{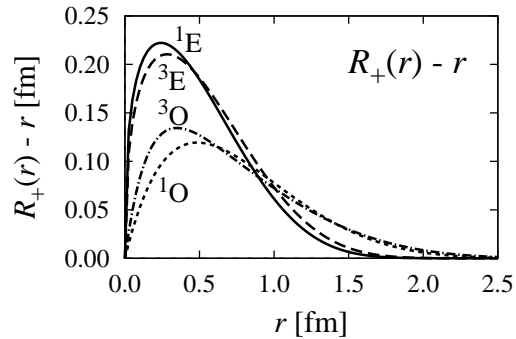}
\caption{The UCOM functions $R_+(r)$ used in even ($^{1,3}E$) and odd ($^{1,3}O$) channels.  The differences between $R_+(r)$ and the original coordinate $r$ are displayed.}
\label{R+}
\end{center}
\end{figure}
%%%%%%%%%%%%%%%%%%%%%%%%%%%%%%

In the framework of UCOM, it is generally possible to introduce the partial wave dependence in $R_+(r)$ and then the $R_+(r)$ functions are determined in each relative partial wave of the nucleon pair.  In the previous study \cite{myo09}, we have performed the extension of UCOM by taking care of the characteristics of the short range correlation.  One of the simplest extensions is UCOM for only $s$-wave relative motion.  The partial waves with finite orbital angular momentum $l$ except for the $s$-wave ($l=0$) have $r^l$ behavior of the relative wave function near the origin. This $r^l$ behavior cuts down the effect of the short-range hard core in the $NN$ interaction largely. On the other hand, only the $s$-wave function is finite at the origin and its behavior near the origin is determined by the hard core dynamics.  Considering this physical characteristics of the $s$-wave, we have introduced $S$-UCOM, in which the UCOM transformation is performed for only the relative $s$-wave component in the even channel.  In the ordinary UCOM, there is too large a removal of the short-range part of the relative wave functions, in particular, in the $d$-wave part used in the $sd$ coupling of the tensor matrix elements, where the tensor interaction possesses some amount of the strength.  In order to overcome this feature, we have newly introduced $S$-UCOM and confirmed that the situation was improved for the binding energy by a few MeV in $^4$He~\cite{myo09}.  In TOSM with $S$-UCOM, we can optimize the $R_+(r)$ functions and the optimized three parameters in $S$-UCOM for $^4$He are listed in Table \ref{tab:R+2}.  The demonstration of the calculation to search for the energy minimum is shown in Ref.~\cite{myo09}.  The adopted $R_+(r)$ in $S$-UCOM for four channels are shown in Fig.~\ref{R+}. In the present analysis, we use these $R_{+}(r)$ functions for every states of He isotopes.  To simplify the numerical calculation, we adopt the ordinary UCOM for the central correlation part instead of the $S$-UCOM in this analysis. 

%%%%%%%%%%%%%%%%%%%%%%%%%%%%%%%%%%%%%%%%%%%%%%%%%%%%%%%%%%%
\section{Results}\label{sec:result}
\subsection{$^4$He}

\begin{figure}[t]
\begin{center}
\includegraphics[width=7.0cm,clip]{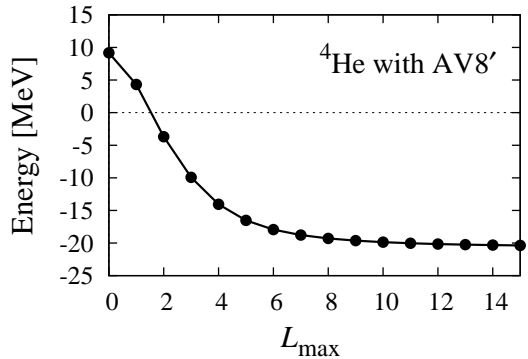}
\caption{Energy of $^4$He in TOSM+UCOM as a function of the maximum orbital angular momentum $L_{\rm max}$.}
\label{ene_4He}
\end{center}
\end{figure}

We show first the results of $^4$He using TOSM+UCOM.  We then explain the structures of $^{5-8}$He and discuss how their structures change in comparison with that of $^4$He.  We explain numerical results of $^4$He using the AV8$^\prime$ interaction, which consists of central, $LS$ and tensor terms and is used in the calculation given by Kamada {\it et al}, where the Coulomb term is ignored \cite{kamada01}.  In Fig. \ref{ene_4He}, the convergence of the energy of $^4$He is shown as a function of $L_{\rm max}$, the maximum orbital angular momentum of particle states in TOSM.  We get a good convergence for $^4$He. The Hamiltonian components are shown in Table \ref{tab:4He_ham} in comparison with the stochastic variational method (SVM) using correlated Gaussian basis functions\cite{varga95,suzuki08}, which is one of the rigorous calculations~\cite{kamada01}. The matter radius of $^4$He is obtained as 1.52 fm in TOSM+UCOM.  When we apply the $S$-UCOM instead of ordinary UCOM, the energy gain is about 2 MeV in total energy and gets closer to the rigorous value \cite{myo09}.  In particular, contribution from the tensor interaction becomes large due to the improvement of the $sd$ coupling of the tensor matrix elements as was mentioned.  

As for the comparison with the rigorous calculation, we see that contribution from the central interaction in TOSM+UCOM/$S$-UCOM satisfies the rigorous value.  On the other hand, the tensor energy and the kinetic energy show some shortage from the rigorous values even in the $S$-UCOM case.  One of the possible explainations for the shortage of those energies in this approach is the treatment of the short-range part of the tensor correlations.  Although the dominant part of the tensor interaction is of intermediate and long ranges, there remains small strength in the short-range part of the tensor interaction, which can couple with the short-range correlations.  
Horii {\it et al.} estimate the amount of this coupling by using the few-body method of SVM \cite{horii11},
in which the short-range repulsion and the tensor interaction are directly treated.
In Ref. \cite{horii11}, they propose the one-pair approximation of tensor coupling in the few-body wave function, named tensor-optimized few-body model (TOFM). It is shown that the TOFM gives the good binding energy of $^4$He of 24.05 MeV with AV8$^\prime$ as compared with the rigorous calculation. The physical concept of TOFM is the same as that of TOSM except for the use of UCOM.  These results imply that the UCOM transformation for short-range correlation makes the energy loss of about 1.5 MeV in $^4$He in comparison with the TOFM result, in particular, for the tensor contribution. The remaining shortage of the binding energy with respect to the rigorous calculation shown in Table \ref{tab:4He_ham} should come from higher configurations beyond $2p2h$ excitations in TOSM.  Horii {\it et al.} also confirmed that the two-pair tensor coupling in $^4$He by extending TOFM, reproduces the rigorous results within 200 keV \cite{horii11-2}, which corresponds to the $4p4h$ mixing in TOSM.

Considering the difference between the results of TOSM and TOFM, three-body term of the UCOM transformation is one of the possibilities to overcome the lack of energy from UCOM~\cite{myo09, feldmeier98}.  This three-body term may contribute to the increase the tensor energy and bring more high momentum components in the wave function.  It is also noted that it has been discussed that the three-body term of UCOM can work repulsively more or less for the Hamiltonian having only the central $NN$ interaction \cite{feldmeier98} and that a similar trend can be seen in UCOM including the tensor-type transformation \cite{roth10}.  Another possibility is the improvement of the correlation function $R_+(r)$.  In this calculation, the functional forms of $R_+(r)$ shown in Eqs. (\ref{eq:R+}) and (\ref{eq:R-}) are introduced from the consideration of the short-range behavior of the two-body system only with the central $NN$ interaction \cite{feldmeier98}.  It would be interesting to study an appropriate form of $R_+(r)$ which is suitable for the Hamiltonian including the tensor interaction explicitly, and discuss the effect of the many-body term of UCOM.  The detailed analysis of the result obtained in TOFM \cite{horii11} gives some ideas to determine the form of $R_+(r)$ from the variational principle.

%%%%%%%%%%%%%%%%%%%%%%%%%%%%%%
%%%%%%%%%%%%%%%%%%%%%%%%%%%%%%
%%%%%%%%%%%%%%%%%%%%%%%%
\begin{table}[t]
\caption{Various energy components in MeV for $^4$He.}
\begin{tabular}{c|ccccc}
\noalign{\hrule height 0.5pt}
               &  Energy  &  Kinetic  &   Central  & Tensor  &  $LS$    \\
\noalign{\hrule height 0.5pt}
TOSM+UCOM          & $-20.46$ & $ 86.95$ & $-54.63$ & $-51.06$  & $-1.73$ \\   
TOSM+$S$-UCOM      & $-22.30$ & $ 90.50$ & $-55.71$ & $-54.55$  & $-2.53$ \\
SVM~\cite{kamada01}& $-25.92$ & $102.35$ & $-55.23$ & $-68.32$  & $-4.71$ \\
\noalign{\hrule height 0.5pt}
\end{tabular}
\label{tab:4He_ham}
\end{table}
%%%%%%%%%%%%%%%%%%%%%%%%
%%%%%%%%%%%%%%%%%%%%%%%%%%%%%%%%%%%%%%%%%%%%%%%%%%%%%%%%%%%%%%%
\begin{table}[t]
\caption{Mixing probabilities in \% for $^4$He with TOSM+UCOM, where the two subscripts $00$ and $10$ are the spin-isospin quantum numbers.}
\begin{tabular}{c|r}
\noalign{\hrule height 0.5pt}  %Src10.6/Data08/
$(0s)_{00}^4$                                  & 84.14  \\
$(0s)_{10}^{-2}(0p_{1/2})_{10}^2$              &  2.32  \\
$(0s)_{10}^{-2}[(1s_{1/2})(0d_{3/2})]_{10}$    &  2.20  \\
$(0s)_{10}^{-2}[(0p_{3/2})(0f_{5/2})]_{10}$    &  1.82  \\
$(0s)_{10}^{-2}[(0p_{1/2})(0p_{3/2})]_{10}$    &  1.21  \\
$(0s)_{10}^{-2}[(0d_{5/2})(0g_{7/2})]_{10}$    &  0.78  \\
\noalign{\hrule height 0.5pt}
\mbox{remaining part}                          &  7.53  \\
\noalign{\hrule height 0.5pt}
\end{tabular}
\label{tab:mixing} 
\end{table}
%%%%%%%%%%%%%%%%%%%%%%%%%%%%%%%%%%%%%%%%%%%%%%%%%%%%%%%%%%%%%%%

%%%%%%%%%%%%%%%%%%%%%%%%%%%%%%
\begin{table}[th]  % TOSM2/Data04/He4_b1.5.tmp
\caption{Occupation numbers in each orbit in $^4$He.}
\begin{tabular}{c|cccccccc}
\noalign{\hrule height 0.5pt}
$^4$He$(J^\pi)$  &~$s_{1/2}$~&~$p_{1/2}$~&~$p_{3/2}$~&~$d_{3/2}$~&~$d_{5/2}$~\\ %  & $f_{5/2}$ & $f_{7/2}$ &  Others  \\
\noalign{\hrule height 0.5pt}
$0^+$            & ~3.80~    &~ 0.06 ~   & ~ 0.04 ~  &~  0.04  ~ &~ 0.01~     \\ %  &  0.022    &   0.006   &   0.022  \\
\noalign{\hrule height 0.5pt}
\end{tabular}
\label{tab:occ_4He}
\end{table}

To see the properties of the $^4$He wave function, dominant configurations are listed in Table \ref{tab:mixing} with their probabilities. For the particle states, the various radial components are summed up for each orbit. It is found that the specific $2p2h$ states such as $(0s)_{10}^{-2}(0p_{1/2})_{10}^2$ show large probabilities and these configurations are essential to produce the tensor correlation in $^4$He because of the coupling by the tensor operator \cite{myo07, toki02}. These dominant $2p2h$ states commonly have the isoscalar nature and correspond to the excitations of a $pn$ pair, which indicates the deuteron-like correlation. It has already been shown that the particle states in those $2p2h$ states have the spatially compact radial wave functions \cite{myo07, sugimoto04, myo05}. It is also found that the $0p0h$-$2p2h$ coupling exhausts 94\% of the tensor energy. Occupation numbers in various nucleon orbits are shown in Table \ref{tab:occ_4He}.  The $p_{1/2}$ orbit has the largest occupation number among particle states because of the large $2p2h$ mixings including the $p_{1/2}$ component.  This feature of $2p2h$ excitations plays an important role to determine the structures of $^{5-8}$He as will be discussed later.

%%%%%%%%%%%%%%%%%%%%%%%%%%%%%%%%%%%%%%%%%%%%%%%%%%%%%%%%%%%%%%%
\subsection{Energy spectra of He isotopes}

\begin{figure}[thb]
\centering
\includegraphics[width=8.5cm,clip]{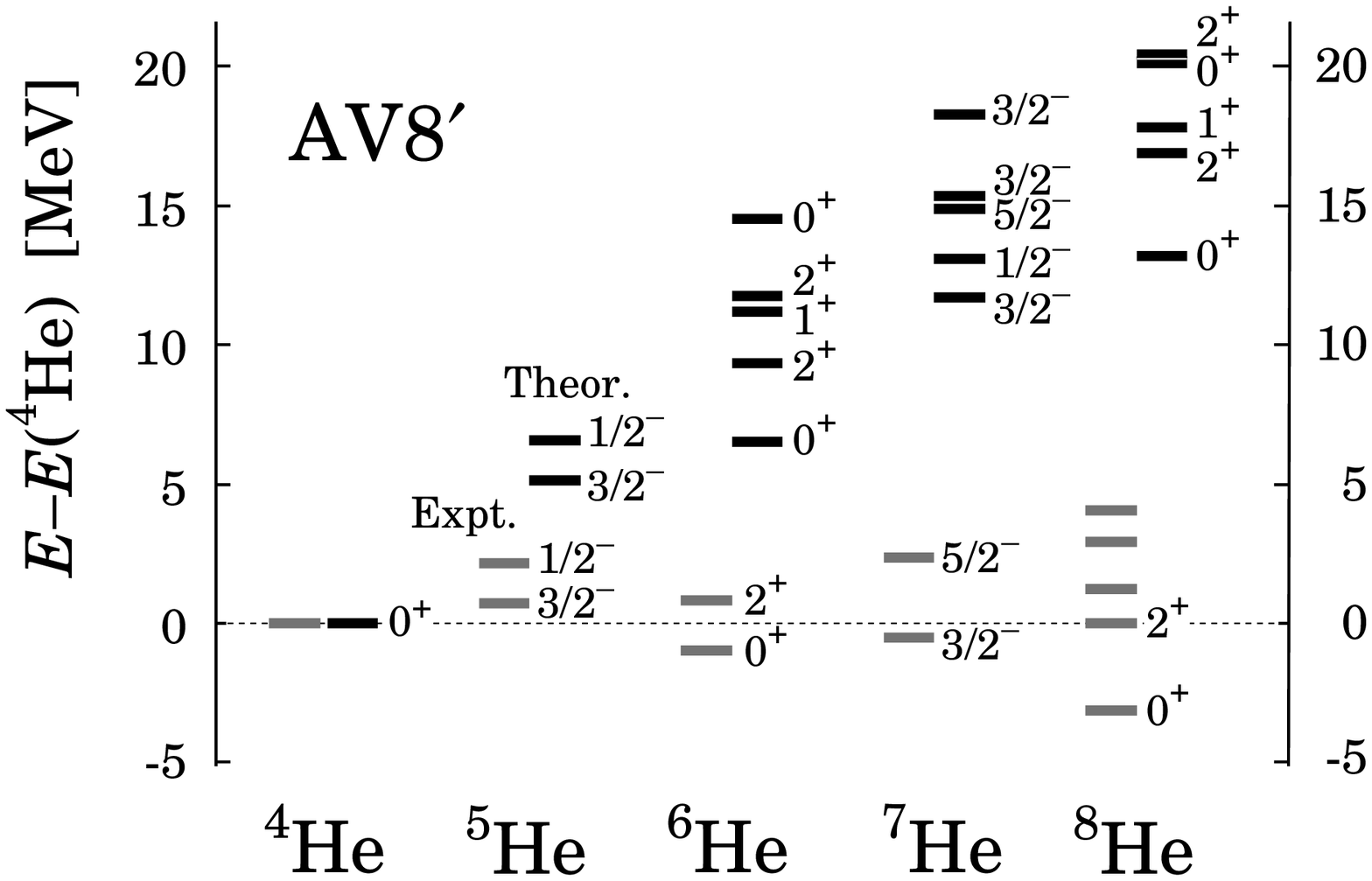}
\caption{Energy levels of He isotopes using AV8$^\prime$, normalized to the $^4$He ground state energy.  Experimental data are taken from Refs. \cite{skaza06, skaza07, golovkov09}.} 
\label{fig:AV8}
\vspace*{0.0cm}
\centering
\includegraphics[width=8.5cm,clip]{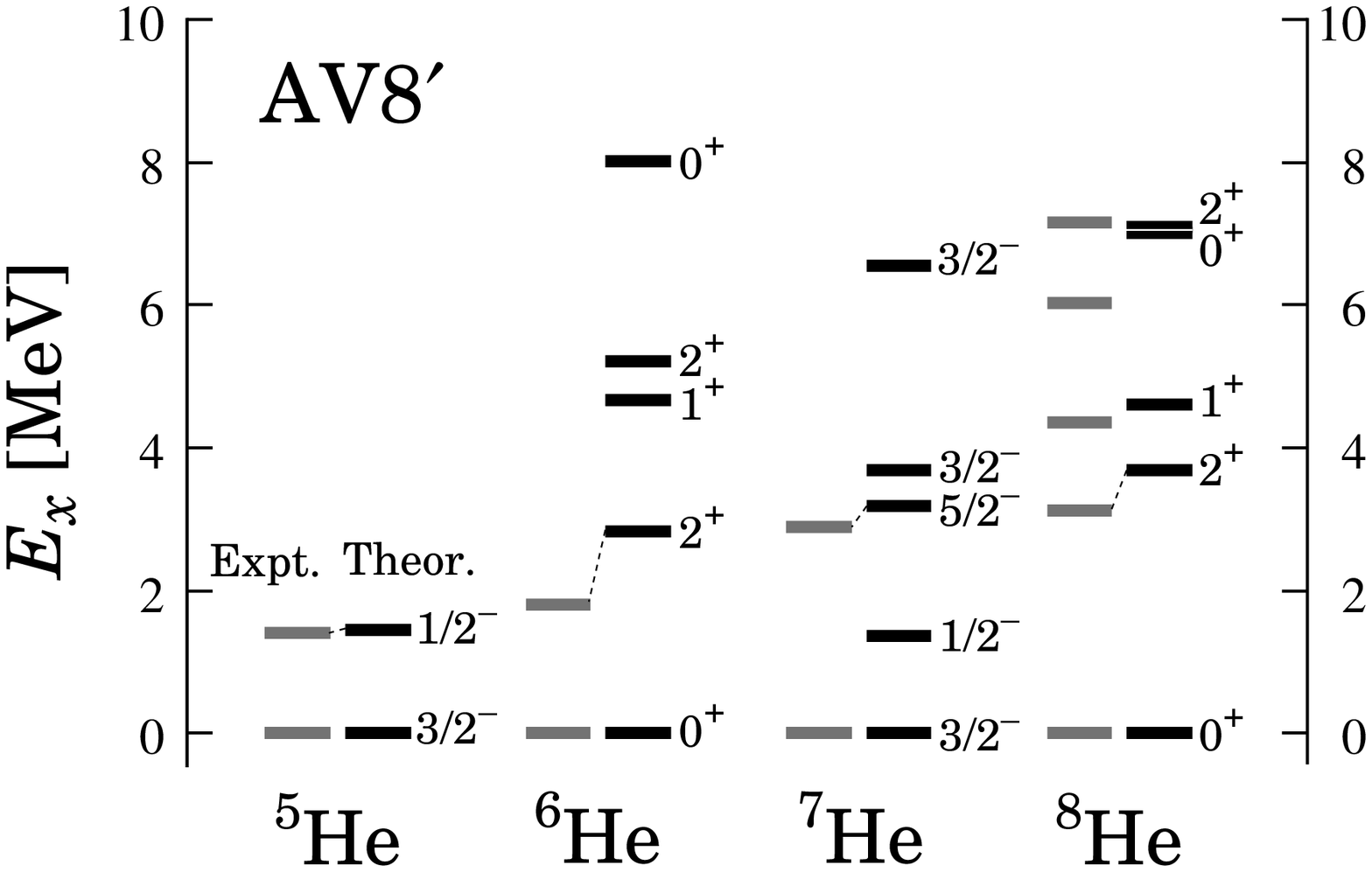}
\caption{Excitation energies of He isotopes using AV8$^\prime$.} 
\label{fig:AV8ex}
\vspace*{0.0cm}
\centering
\includegraphics[width=8.5cm,clip]{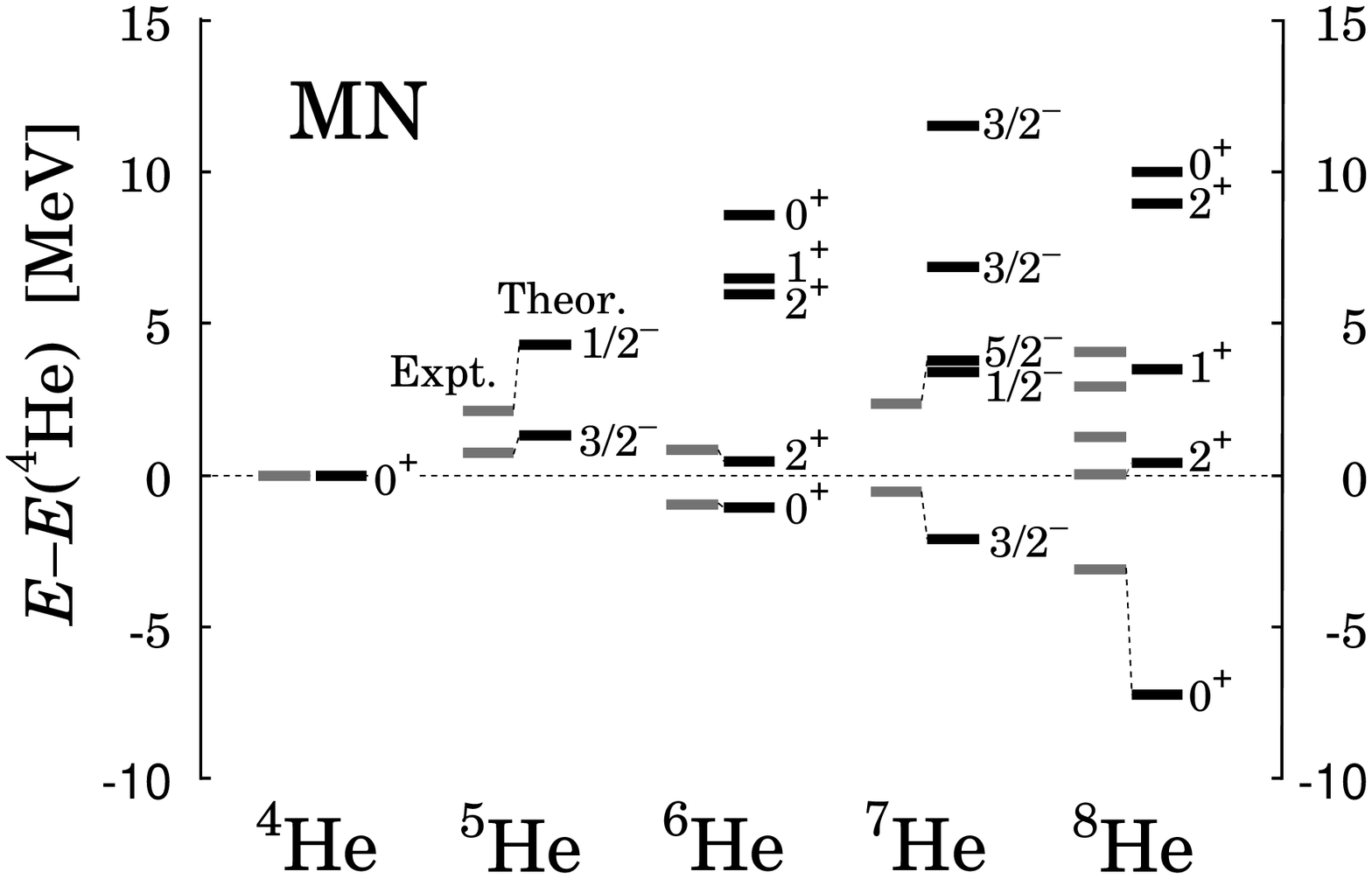}
\caption{Energy levels of the He isotopes using the Minnesota interaction normalized to the $^4$He ground state energy.}
\label{fig:MN}
\end{figure}

We explain the results of neutron-rich He isotopes using TOSM+UCOM with the AV8$^\prime$ interaction, where $L_{\rm max} $ is taken as 10 to get a sufficient convergence. We show the energy spectra with respect to the $^4$He ground state energy in Fig. \ref{fig:AV8}. The excitation energies of $^{5-8}$He are shown in Fig. \ref{fig:AV8ex}.  The matter radii of $^6$He and $^8$He are listed in Table \ref{tab:radius},
which are slightly smaller than those of the experiments \cite{tanihata92, alkazov97, kiselev05}.  We also show the results of cluster orbital shell model (COSM) \cite{myo10}, in which the spatial extension of extra neutrons are fully described. This point will be discussed later.
For charge radius, the precise data of $^{6,8}$He are reported \cite{mueller07}.
To obtain the charge radius it is necessary to calculate the isospin-breaking matrix elements, which are not yet obtained in the present TOSM.

%%%%%%%%%%%%%%%%%%%%%%%%%%%%%%                                                                                                                                                     
\begin{table}[t]
\caption{Matter radii of $^6$He and $^8$He in comparison with the cluster orbital shell model (COSM) \cite{myo10} and 
the experiments; a\cite{tanihata92}, b\cite{alkazov97}, c\cite{kiselev05}. Units are in fm.}
\label{tab:radius}
\centering
\begin{tabular}{r|p{1.3cm} p{1.3cm} p{4.0cm}}
\hline
       & Present  & COSM   & Experiment        \\
\hline
$^6$He    &~~2.27 &~~2.37  & 2.33(4)$^{\rm a}$~~~~2.45(10)$^{\rm b}$~~~~2.37(5)$^{\rm c}$ \\
$^8$He    &~~2.44 &~~2.52  & 2.49(4)$^{\rm a}$~~~~2.53(8)$^{\rm b}$~~~~2.49(4)$^{\rm c}$ \\
\hline
\end{tabular}
\end{table}
%%%%%%%%%%%%%%%%%%%%%%%%%%%%%%                                                                                                                                                      

It is found that our results underestimate the binding energies of He isotopes and its amount becomes larger for neutron-rich side.  
On the other hand, for excitation energies shown in Fig. \ref{fig:AV8ex}, we see a good correspondence to the experimental energy levels.  This result indicates that the level spacing is described well in TOSM+UCOM, so that we can discuss the structure differences between energy levels.  Before doing this, we discuss several reasons of under-binding, which can contribute to the bulk part of the binding energies, such as the continuum effect of extra neutrons having a weak binding nature, higher configurations beyond $2p2h$ states in TOSM, the genuine three-body interaction, and the improvement of UCOM.

As for the continuum effect, the spatial extension of extra neutrons is explicitly treated in the TOSM basis within the $2p2h$ excitations using the Gaussian expansion method by taking long-range parameters in Eq. (\ref{eq:Gauss2}).  However, the continuum effect of extra neutrons is not sufficiently described in the TOSM approximation.   When the $2p2h$ excitations in TOSM are used to describe the $pn$ tensor correlation in $^4$He, the extra neutrons still occupy the $p$-orbits in the hole states.  This means the continuum effect is not taken into account together with the large energy gain due to the tensor correlation in TOSM.  Hence, we estimate the correlation energy from the continuum effect by using COSM \cite{myo10}, in which each motion of extra neutrons is solved exactly around the core with no tensor correlation, and the continuum effect of multi-neutrons is fully included.  It is obtained that the correlation energies beyond the $p$-shell configuration of extra neutrons are estimated as 3.3 MeV for $^6$He and 6.6 MeV for $^8$He in their ground states, respectively \cite{myo10}.  

It is important then to estimate how many particle-many hole configurations are necessary to include the continuum effect of extra neutrons.  For $^8$He, when we limit COSM up to $2p2h$ excitations of extra four neutrons, the energy gain is 5.9 MeV, which is about 90\% of the correlation energy of the continuum effect of 6.6 MeV. This result indicates that a large fraction of the continuum effect can be described within the $2p2h$ space from the $p$-shell neutrons.  This fact suggests that the extension of TOSM to include $4p4h$ configurations as a whole can explain this feature of extra neutrons for He isotopes with two-kinds of $2p2h$ contributions; 
a $pn$ pair with $T=0$ in the $^4$He core and a $nn$ pair with $T=1$ in the $p$-shell neutrons.  Furthermore, to satisfy the boundary condition of neutron emissions, it is necessary to connect the TOSM basis to the correct asymptotic wave functions such as the $^4$He+$n$+$n$ three-body state for $^6$He \cite{myo10}.  This is important to describe the tail behavior of the neutron wave functions.

From the GFMC calculation \cite{pieper01}, we can simply estimate the effect of the genuine three-body interaction on the binding energy which increases with the mass number.  For $^8$He, for example, the relative energy gain from three-body interaction is about 5.5 MeV as compared with $^4$He.
This result indicates that for $^8$He, the total energy gain of about 12 MeV is expected considering two effects as the continuum states and the genuine three-body interaction. Hence, the remaining missing energy from the experimental value is estimated to be as small as 3-4 MeV for $^8$He from Fig. \ref{fig:AV8}.

For the UCOM part, we are able to increase the tensor contributions and the binding energies by using $S$-UCOM, which can improve short-range part of the $sd$ coupling of the tensor interaction.
The improvement of the UCOM correlation function $R_+(r)$ provides another possible way to increase the energy, according to the results obtained in TOFM \cite{horii11} as explained above.

In He isotopes, most of the states are observed as resonances and in this analysis we describe those states within the bound state approximation.   This treatment makes it possible to focus the discussion on the internal structures of each level in He isotopes systematically.  In the previous studies \cite{myo05, myo06, myo07_11}, we include the continuum effects explicitly, using the TOSM wave function as the core nucleus in the extended two- and three-body cluster models.  In those analyses, we have successfully explained the $^4$He+$n$ scattering phase shifts, neutron halo formation and three-body Coulomb breakup strengths in halo nuclei $^6$He and $^{11}$Li and so on, although the systems are not described starting from the single $NN$ interaction like the present TOSM+UCOM.
It was found that the $2p2h$ components induced by the tensor interaction play an important role in reproducing the physical observables in those studies. 
Furthermore, the resulting characteristics of the tensor interaction are essentially the same as that obtained in the present analysis.
Based on these achievements, in the present study it is important to understand the role of tensor interaction in each state of He isotopes systematically fully using TOSM+UCOM starting from the bare $NN$ interaction.  The present approach can also be extended to the analysis which includes the continuum states explicitly, such as the scattering state in the $^4$He+$n$ system \cite{nollett07}.

%%%%%%%%%%%%%%%%%%%%%%%%%%%%%
%%%%%%%%%%%%%%%%%%%%%%%%%%%%%%

To see the interaction dependence, the energy spectra using the Minnesota (MN) interaction are shown in Fig. \ref{fig:MN}. The MN two-body interaction consists of the central and $LS$ parts with $u$ parameter being 0.95 \cite{reichstein70, tang78}.  We use Set III of the MN interaction corresponding to $u=0.95$ for the $LS$ force\cite{reichstein70}.  There is no explicit tensor correlation in the MN interaction. We do not use UCOM for the calculation using the MN interaction, because the MN interaction does not have a strong short-range repulsion.  It was already confirmed that TOSM can reproduce the binding energy and the radius of $^4$He obtained by few-body SVM approach with a central interaction case~\cite{varga95}. 
The difference of binding energy with MN is about 200 keV in comparison with the SVM approach \cite{varga95}.
Using the MN interaction, the binding energy of $^4$He is obtained as 30.55 MeV without the Coulomb interaction. The radius of $^4$He is obtained as 1.39 fm, which is smaller than the experiment by about 0.1 fm. 
For $^6$He and $^8$He, their matter radii are obtained as 1.93 fm and 1.91 fm, respectively, which are smaller than the experiments \cite{tanihata92, alkazov97, kiselev05} shown in Table \ref{tab:radius} by about 0.5 fm.
From the energy spectra shown in Fig. \ref{fig:MN}, it is found that the reproduction of $^6$He energy is good, on the other hand the energies of $^7$He and $^8$He are overestimated. One of the reasons is the too large splitting energy between $p_{3/2}$ and $p_{1/2}$ components mainly given by the $LS$ interaction. When we adjust the MN interaction such as changing the $u$ parameter and the strength of the $LS$ interaction, it is still difficult to reproduce the whole trend of the energies of the ground and excited states of He isotopes consistently.  One of the reasons to explain this tendency is that the MN interaction is originally formulated as an effective interaction when $^4$He is assumed to be the $(0s)^4$ configuration~\cite{tang78, varga94, arai99, myo01}.  This condition is not imposed in TOSM.  Another difficulty of the MN interaction is the radius property, which is related to the nuclear saturation density. The MN interaction is known to give spatially compact wave functions than the experimental values, even for the $(0s)^4$ state of $^4$He~\cite{varga94}.

%%%%%%%%%%%%%%%%%%%%%%%%%%%%%%
\subsection{Hamiltonian components}

We discuss the structures of each level of He isotopes.
Various energy components in $^{5-8}$He are listed in Tables \ref{tab:5He}, \ref{tab:6He}, \ref{tab:7He} and \ref{tab:8He},
which are useful to discuss the structures of individual states.
In this calculation, we take the common length parameters of the hole states, $0s$ and $0p$ orbits. This is done to exclude the continuum effect, which produces a few MeV energy gain in $^5$He as shown in Fig~\ref{fig:AV8}, and to focus our discussion on the internal structures of He isotopes.  We take 1.5 fm of the length parameter of the hole states, which is the optimized value of the $0s$ orbit in $^4$He. This length parameter corresponds to $\hbar\omega$ as 18.43 MeV.
In the full calculations explained above, we minimize the energies by optimizing the length parameters of all the hole states for each $J^\pi$ state of He isotopes.

%%%%%%%%%%%%%%%%%%%%%%%%%%%%%%%%%%%%%%%%%%%%%%%%%%%%%%
\begin{table}[t]
\caption{Various energy components in $^5$He measured from those of the $^4$He ground state. Energy units are given in MeV and $\hbar\omega=18.43$ MeV.  Dominant configurations of the extra neutron are shown also.}
\begin{tabular}{ccccccc}
\noalign{\hrule height 0.5pt}
$^5$He$(J^\pi)$ & Config.   & Energy  & Kinetic & Central & Tensor  & $LS$     \\
\noalign{\hrule height 0.5pt}
3/2$^-$         & $p_{3/2}$ & $6.97$  & $24.14$ & $-8.99$ & $-5.60$ & $-2.58$  \\
1/2$^-$         & $p_{1/2}$ & $10.05$ & $17.53$ & $-6.96$ & $-1.11$ & $1.04$   \\
\noalign{\hrule height 0.5pt}
\end{tabular}
\label{tab:5He}
%%%%%%%%%%%%%%%%%%%%%%%%%%%%%%%%%%%%%%%%%%%%%%%%%%%%%%%
\caption{Various energy components in $^6$He measured from those of the $^4$He ground state. Units are in MeV.}
\begin{tabular}{ccccccc}
\noalign{\hrule height 0.5pt}
$^6$He$(J^\pi)$ & Config.              & Energy  & Kinetic & Central  & Tensor   & $LS$     \\
\noalign{\hrule height 0.5pt}
$0^+_1$         & $(p_{3/2})^2$        &  $ 8.95$ & $53.04$ & $-27.75$ & $-12.02$ & $-4.04$  \\
$0^+_2$         & $(p_{1/2})^2$        &  $21.90$ & $34.30$ & $-14.06$ & $ -0.17$ & $ 2.11$  \\ 
$1^+$           & $(p_{3/2})(p_{1/2})$ &  $17.29$ & $42.90$ & $-15.98$ & $ -8.49$ & $-0.86$  \\
$2^+_1$         & $(p_{3/2})^2$        &  $11.40$ & $52.41$ & $-22.93$ & $-12.80$ & $-4.99$  \\
$2^+_2$         & $(p_{3/2})(p_{1/2})$  &  $16.07$ & $45.06$ & $-18.06$ & $ -8.54$ & $-2.10$  \\
\noalign{\hrule height 0.5pt}
\end{tabular}
\label{tab:6He}
\end{table}

\begin{figure}[tb]
\centering
\includegraphics[width=6.7cm,clip]{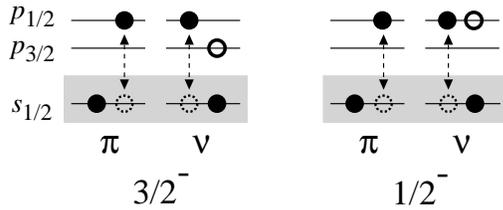}
\caption{Configurations in $^5$He. Open circles with solid lines indicates the extra neutrons.}
\label{fig:config_He5}
\end{figure}

In Table \ref{tab:5He}, we compare various energy components in $3/2^-$ and $1/2^-$ states of $^5$He measured from those of $^4$He. The $LS$ splitting energy obtained is about 3 MeV. We discuss the effect of the tensor interaction on this splitting energy between the two states. A large difference is seen in the tensor energy of the $3/2^-$ state as compared with that of $1/2^-$. The larger contribution of the tensor interaction in $3/2^-$ brings the enhancement of the kinetic energy because of the involvement of high momentum components from the tensor interaction.  The amount of the enhanced kinetic energy is 24 MeV, which is larger than $\hbar\omega$.   For $1/2^-$, on the other hand, the energy gain from the tensor interaction is small and the enhancement of the kinetic energy is 17.5 MeV, which is close to the value of $\hbar\omega$. These results are related to the larger amount of the $p_{1/2}$ component in $^4$He than the $p_{3/2}$ one as shown in Table \ref{tab:mixing}.  This situation is explained in Fig.~\ref{fig:config_He5},  where the last neutron in $^5$He occupies the $p_{3/2}$ orbit in the left panel, this neutron does not disturb the $^4$He structure.  Hence, the $p_{3/2}$ occupied state gains an additional tensor energy without disturbing the large energy gain in $^4$He.  On the other hand, in case of the dominant $p_{1/2}$ occupation for the $1/2^-$ state shown in the right panel of Fig.~\ref{fig:config_He5}, this neutron blocks some component of the spatially compact $p_{1/2}$ neutron in the $2p2h$ excitations of the $^4$He core configuration because of the small degeneracy of the $p_{1/2}$ orbit. This effect dynamically produces the Pauli-blocking and reduces the total binding energy of $^5$He.  As a result, the last neutron located in the $p_{1/2}$ orbit should be orthogonal to the excited $p_{1/2}$ orbit in $^4$He and the tensor interaction does not gain the energy in $^5$He($1/2^-$). The coupling behavior between $^4$He and a last neutron explains the difference in the Hamiltonian components in two states of $^5$He, which results in the $LS$ splitting energy as a net value. 

%%%%%%%%%%%%%%%%%%%%%%%%%%%%%
%%%%%%%%%%%%%%%%%%%%%%%%%%%%%

%%%%%%%%%%%%%%%%%%%%%%%%%%%%%
\begin{figure}[tb]
\centering
\includegraphics[width=7.5cm,clip]{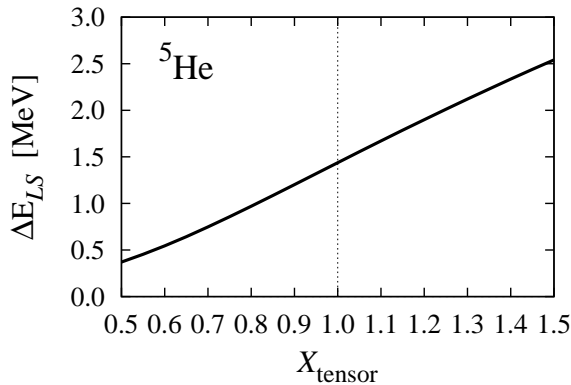}
\caption{$LS$ splitting energy in $^5$He as a function of the strength of the tensor interaction $X_{\rm tensor}$, which is multiplied to the tensor interaction in AV8$^\prime$.}
\label{fig:5He_LS}
\end{figure}
%%%%%%%%%%%%%%%%%%%%%%%%%%%%%%

In order to show directly the relation between the $LS$ splitting energy in $^5$He and the tensor interaction, we show the $LS$ splitting energy of $^5$He between $3/2^-$ and $1/2^-$ states in Fig.~\ref{fig:5He_LS} by changing the strength of tensor interaction. In this process, we always optimize the length parameters of hole states independently in the same manner as is done in Fig. \ref{fig:AV8}.  When the tensor interaction is enhanced, the $LS$ splitting energy becomes larger, because of the different couplings between $^4$He and a last neutron occupied in the $p_{1/2}$ or $p_{3/2}$ orbit.  This result shows that the tensor interaction plays a decisive role to create the $LS$ splitting energy in $^5$He \cite{terasawa59,arima60,nagata59}. In our previous studies\cite{myo05,myo06}, we have estimated the contribution of this tensor correlation to the observed phase shifts of the $^4$He-$n$ scattering, which produces about 30\% of the splitting. A similar tendency is confirmed for other He isotopes, such as the level spacing between the $0^+_1$ and $0^+_2$ states in $^6$He.  

\begin{figure}[tb]
\centering
\includegraphics[width=6.7cm,clip]{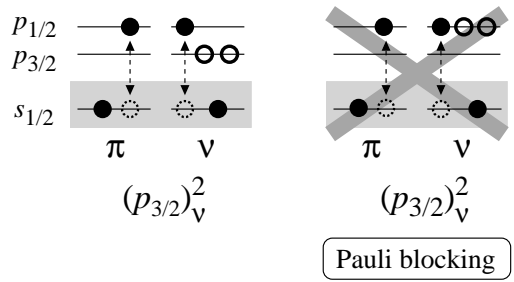}
\caption{Configurations in $^6$He. Open circles with solid lines indicates the extra neutrons.}
\label{fig:config_He6}
\end{figure}

For $^6$He, various energy components are shown in Table \ref{tab:6He}, similarly to $^5$He.  We can classify the obtained five states of $^6$He into three groups by seeing the tensor energies. The first group is the ground $0^+_1$ and $2^+_1$ states, where the extra two neutrons dominantly occupy the $(p_{3/2})^2$ configuration. In this case the tensor correlation in $^4$He is hardly disturbed due to a small mixing of the $p_{3/2}$ neutron in $^4$He, thus the tensor interaction acts attractively to the binding energy as illustrated in the left panel of Fig. \ref{fig:config_He6}. The difference of the central interaction energies in two states indicates the existence of the neutron pairing correlation in the $0^+_1$ state. In these states, we see also the enhancement of the kinetic energies coming from the high momentum component associated with the tensor interaction beyond the value of $2\times \hbar\omega$. The second group is the $1^+$ and $2^+_2$ states, where the extra two neutrons occupy the $(p_{1/2}\; p_{3/2})$ configuration.  In these states, similarly to the $^5$He($1/2^-$) case, tensor energy and kinetic energy decrease from those in the first group because one of the extra neutrons occupies the $p_{1/2}$ orbit and the blocking of this orbit occurs. The third group is the $0^+_2$ state with the main occupation probability in the $(p_{1/2})^2$ configuration for the extra two neutrons.  In this state, the Pauli-blocking effect to block the tensor correlation strongly occurs as shown in the right panel of Fig.~\ref{fig:config_He6}.  As a result, the tensor interaction cannot produce the energy gain from the $^4$He case and the kinetic energy shows the lowest value among the $^6$He states. This blocking dynamically produces the $LS$ splitting energy and explains the level spacing in $^6$He.  From these results of $^6$He, it is found that the configurations of extra two neutrons in $^6$He are essential to explain the difference seen in various Hamiltonian components of the individual state shown in Table \ref{tab:6He}. Enhancement of the kinetic energy as the increase of the high momentum component beyond the $\hbar\omega$ effect is also originated from the enhancement of the tensor energy.

%%%%%%%%%%%%%%%%%%%%%%%%%%%%%
%%%%%%%%%%%%%%%%%%%%%%%%%%%%%%

%%%%%%%%%%%%%%%%%%%%%%%%%%%%%%%%%%%%%%%%%%%%%%%%%%%%%%%
\begin{table}[t]
\caption{Various energy components in $^7$He measured from those of the $^4$He ground state. Units are in MeV.}
\begin{tabular}{ccccccc}
\noalign{\hrule height 0.5pt}
$^7$He$(J^\pi)$ & Config.                  & Energy & Kinetic& Central  & Tensor    & $LS$     \\
\noalign{\hrule height 0.5pt}
$3/2^-_1$       & $(p_{3/2})^3$            & 14.97  & 82.89  & $-39.13$ & $ -19.89$ & $ -8.63$ \\
$3/2^-_2$       & $(p_{3/2})^2_2(p_{1/2})$ & 21.44  & 74.14  & $-33.64$ & $ -15.98$ & $ -2.80$ \\ 
$3/2^-_3$       & $(p_{3/2})^3(p_{1/2})^2$ & 27.50  & 65.82  & $-27.09$ & $  -9.96$ & $ -0.98$ \\ 
$1/2^-$         & $(p_{3/2})^2_0(p_{1/2})$ & 19.07  & 76.81  & $-38.52$ & $ -15.74$ & $ -3.20$ \\
$5/2^-$         & $(p_{3/2})^2_2(p_{1/2})$ & 19.92  & 77.53  & $-36.31$ & $ -16.60$ & $ -4.37$ \\
\noalign{\hrule height 0.5pt}
\end{tabular}
\label{tab:7He}
%\end{table}
%%%%%%%%%%%%%%%%%%%%%%%%%%%%%%%%%%%%%%%%%%%%%%%%%%%%%%%
%\begin{table}[t]
\caption{Various energy components in $^8$He measured from those of the $^4$He ground state. Units are in MeV.}
\begin{tabular}{ccccccc}
\noalign{\hrule height 0.5pt}
$^8$He$(J^\pi)$  & Config.                  & Energy & Kinetic& Central  & Tensor   & $LS$     \\
\noalign{\hrule height 0.5pt}
$0^+_1$          & $(p_{3/2})^4$            & 17.57 & 114.87 & $-59.26$ & $-25.97$ & $-11.79$ \\
$0^+_2$          & $(p_{3/2})^2(p_{1/2})^2$ & 29.50 & 101.03 & $-51.60$ & $-17.10$ & $ -2.51$ \\ 
$1^+$            & $(p_{3/2})^3(p_{1/2})  $ & 25.08 & 106.52 & $-50.29$ & $-23.75$ & $ -7.11$ \\
$2^+_1$          & $(p_{3/2})^3(p_{1/2})  $ & 22.54 & 109.60 & $-56.22$ & $-23.42$ & $ -7.13$ \\
$2^+_2$          & $(p_{3/2})^2(p_{1/2})^2$ & 30.13 & 100.15 & $-49.27$ & $-18.08$ & $ -2.38$ \\
\noalign{\hrule height 0.5pt}
\end{tabular}
\label{tab:8He}
\end{table}
%%%%%%%%%%%%%%%%%%%%%%%%%%%%%%%%%%%%%%%%%%%%%%%%%%%%%%%

The similar effect about the roles of tensor correlation as seen in $^{5}$He and $^{6}$He is obtained for $^7$He and $^8$He. For $^7$He, there exist three kinds of groups. The first group, the ground $3/2^-_1$ state mostly exhausts the tensor energy due to the dominant $(p_{3/2})^3$ configuration of the extra three neutrons.  The kinetic energy also shows the large value beyond $3\times \hbar\omega$. The second group, the $3/2^-_2$, $1/2^-$ and $5/2^-$ states show the similar energy components because they have commonly the main configuration of $(p_{3/2})^2(p_{1/2})$ of the extra three neutrons. The third group, the $3/2^-_3$ state shows the smallest tensor and kinetic energies among the $^7$He states. This state mostly has the $p_{1/2}$ component in $^7$He.  For $^8$He, five states are obtained and the three kinds of groups on the contributions of tensor interaction are seen.  The ground $0^+_1$ state has mainly the $(p_{3/2})^4$ configuration of extra four neutrons. Due to this configuration, the $0^+_1$ state mostly exhausts the tensor energy and simultaneously a large kinetic energy beyond that of $4\times \hbar\omega$.  The $1^+$ and $2^+_1$ states form the second group having mainly three $p_{3/2}$ orbit neutrons.  The $0^+_2$ and $2^+_2$ states show small tensor and kinetic energies due to the large $p_{1/2}$ component.

As conclusions of the roles of the tensor interaction in He isotopes, the deuteron-like $pn$ pair of the tensor correlation in $^4$He is coupled with the motion of extra neutrons in neutron-rich He isotopes. There are two-kinds of couplings. One is when the extra neutrons occupy the $p_{3/2}$ orbit, these neutrons do not disturb the tensor correlation in $^4$He due to the small occupation of the $p_{3/2}$ neutron orbit in $^4$He.  As a result, the total system can enhance the tensor contribution and simultaneously the high momentum component induced by the tensor interaction.  On the other hand, when the extra neutrons occupy the $p_{1/2}$ orbit, these neutrons disturb the tensor correlation in $^4$He in some amount, because of the Pauli-blocking effect between the extra $p_{1/2}$ orbit neutron and the $p_{1/2}$ component largely excited from the $s$-orbit in $^4$He by the tensor interaction.  The extra neutrons cannot share the spatially compact $p_{1/2}$ orbit, which is used in $^4$He to produce the $2p2h$ excitations.  As a result, the total system cannot produce a large enhancement of the tensor energy.   Therefore, the contributions of tensor interaction in each level of He isotopes depends on the neutron configurations. This state dependence works to produce the observed $LS$ splitting energy.
The amount of the high momentum component from the tensor interaction also depends on the neutron configurations and explains the kinetic energies of each level.

%%%%%%%%%%%%%%%%%%%%%%%%%%%%%%%%%%%%%%%%%%
\section{Summary}\label{sec:summary}

We have developed a method to describe nuclei with bare $NN$ interaction on the basis of the tensor optimized shell model with the unitary correlation operator method, TOSM+UCOM. We have treated the tensor interaction in terms of TOSM, in which the $2p2h$ states are fully optimized to describe the deuteron-like tensor correlation. The short-range repulsion in the $NN$ interaction is treated in the central correlation part of UCOM. We have shown the reliability of TOSM+UCOM using the AV8$^\prime$ interaction to investigate the structures of He isotopes. It is found that the excitation energy spectra are nicely reproduced.  

It is found that $^4$He contains a relatively large amount of the $pn$ pair in the $p_{1/2}$ orbit. The large mixing of $p_{1/2}$ orbit in $^4$He contributes to the production of the splitting energy between the $LS$ partners in neutron-rich He isotopes.  In $^5$He, it is found that the $3/2^-$ state gains more of the tensor energy than the $1/2^-$ case. This is due to the Pauli-blocking effect between the tensor correlation in $^4$He and the motion of the last neutron occupying the $p_{1/2}$ orbit.  In the $3/2^-$ state of $^5$He, the enhancement of the kinetic energy is observed because of the high momentum component brought by the tensor interaction. As a result, tensor interaction dynamically produces the state dependence in $^5$He and contributes to the creation of the splitting energy between the $1/2^-$ and $3/2^-$ states in $^5$He. The similar mechanism is found in $^{6-8}$He between the configurations seen in the ground and the excited states.  The configurations involving manly the $p_{3/2}$ orbit of extra neutrons gains the tensor energies and produces the large kinetic energies, which are located near the ground state region in He isotopes.  On the other hand, the states involving large $p_{1/2}$ components of extra neutrons are mostly located in the excited states of He isotopes.  In those excited states, the mixing of the $p_{1/2}$ orbit neutrons reduces the enhancement of the tensor energies.  We note that this Pauli-blocking effect of extra neutrons with those states in $2p2h$ excitations due to the strong tensor interaction does not exist for the case of the Minnesota interaction.

The enhancement of the kinetic energy due to the presence of the tensor interaction indicates the increase of the high momentum component in the wave function. Observation of the high momentum component experimentally in finite nuclei in order to confirm the existence of the strong tensor correlation is desired \cite{subedi08, tanihata10}.  Based on the results obtained in TOSM+UCOM for He isotopes, it is interesting to include the continuum component of extra neutrons \cite{myo10} and the genuine three-body interaction.

%%%%%%%%%%%%%%%%%%%%%%
\section*{Acknowledgments}
We thank Professor Hisashi Horiuchi and Professor Kiyoshi Kat\=o for fruitful discussions and continuous encouragement.
This work was supported by a Grant-in-Aid for Young Scientists from the Japan Society for the Promotion of Science (No. 21740194).  Numerical calculations were performed on a computer system at RCNP, Osaka University.

\def\JL#1#2#3#4{ {{\rm #1}} \textbf{#2}, #4 (#3)}  % Physical Review
\nc{\PR}[3]     {\JL{Phys. Rev.}{#1}{#2}{#3}}
\nc{\PRC}[3]    {\JL{Phys. Rev.~C}{#1}{#2}{#3}}
\nc{\PRA}[3]    {\JL{Phys. Rev.~A}{#1}{#2}{#3}}
\nc{\PRL}[3]    {\JL{Phys. Rev. Lett.}{#1}{#2}{#3}}
\nc{\NP}[3]     {\JL{Nucl. Phys.}{#1}{#2}{#3}}
\nc{\NPA}[3]    {\JL{Nucl. Phys.}{A#1}{#2}{#3}}
\nc{\PL}[3]     {\JL{Phys. Lett.}{#1}{#2}{#3}}
\nc{\PLB}[3]    {\JL{Phys. Lett.~B}{#1}{#2}{#3}}
\nc{\PTP}[3]    {\JL{Prog. Theor. Phys.}{#1}{#2}{#3}}
\nc{\PTPS}[3]   {\JL{Prog. Theor. Phys. Suppl.}{#1}{#2}{#3}}
\nc{\PRep}[3]   {\JL{Phys. Rep.}{#1}{#2}{#3}}
\nc{\AP}[3]     {\JL{Ann. Phys.}{#1}{#2}{#3}}
\nc{\JP}[3]     {\JL{J. of Phys.}{#1}{#2}{#3}}
\nc{\andvol}[3] {{\it ibid.}\JL{}{#1}{#2}{#3}}
\nc{\PPNP}[3]   {\JL{Prog. Part. Nucl. Phys.}{#1}{#2}{#3}}
\nc{\FBS}[3]   {\JL{Few Body Syst.}{#1}{#2}{#3}}
%%%%%%%%%%%%%%%%%%%%%%%%%%%%%%%%%%%%%%%%%%%%%%%%%%%%%%%%%%%%%
%\section*{References}

\end{document}